\documentclass[manuscript]{aastex}

\usepackage{subfigure,epsfig,url}
\shorttitle{Decreasing Computing Time with Symplectic Correctors}
\shortauthors{Kaib et al.}

\begin{document}

\title{Decreasing Computing Time with Symplectic Correctors in Adaptive Timestepping Routines}

\author{Nathan A. Kaib\altaffilmark{1}, Thomas Quinn\altaffilmark{1}, and Ramon Brasser\altaffilmark{2}}
\altaffiltext{1}{Department of Astronomy, University of Washington, Box 351580, Seattle, WA 98195-1580, kaib@astro.washington.edu.}
\altaffiltext{2}{Observatoire De La Cote D'Azur, B.P. 4229, 06304 Nice Cedex 4, France}

\begin{abstract}
It has previously been shown that varying the numerical timestep during a symplectic orbital integration leads to a random walk in energy and angular momentum, destroying the phase space-conserving property of symplectic integrators.  Here we show that when altering the timestep symplectic correctors can be used to reduce this error to a negligible level.  Furthermore, these correctors can also be employed to avoid a large error introduction when changing the Hamiltonian's partitioning.  We have constructed a numerical integrator using this technique that is nearly as accurate as widely used fixed-step routines.  In addition, our algorithm is drastically faster for integrations of highly eccentricitic, large semimajor axis orbits, such as those found in the Oort Cloud.
\end{abstract}


\section{Background}

In the past twenty years, mixed variable symplectic integrators have revolutionized dynamical modeling in planetary astronomy.  These integrators work by splitting a system's Hamiltonian into two separate Hamiltonians, $H_{Kep}$ and $H_{Int}$ \citep{wishol91,sahtre92}.  $H_{Kep}$ is the Hamiltonian for a Keplerian orbit about a large central point mass, while $H_{Int}$, or the interaction Hamiltonian, is the difference between the real Hamiltonian and $H_{Kep}$.  These two Hamiltonians are then integrated in a leapfrog algorithm, where a particle is first evolved under the equations of motion of $H_{Kep}$ for a half time step ($\tau$/2), then under the equations of motion of $H_{Int}$ for a full time step ($\tau$), and finally under $H_{Kep}$ again for $\tau$/2.  It can be shown that this is a second-order approximation of the evolution of the real Hamiltonian for a full step, $\tau$ \citep{sahtre92,yosh93}.  (It should be noted that the integration kernel in many popular simulation codes actually integrates $H_{Int}$ for half-steps and $H_{Kep}$ for a full step.  This is also a second order symplectic algorithm.)

Mixed variable symplectic methods have two major advantages over previous integration techniques.  First, symplectic methods can enable much larger integration step sizes compared to other integration techniques if the system being modeled is nearly integrable (in the case of planetary systems, if $H_{Kep}>>H_{Int}$).  For most simulations of a planetary system, $H_{Kep}$ is a fairly close approximation to the real Hamiltonian of the system.  As a result, $H_{Int}$ can be viewed as a perturbation, $\epsilon$, to the Kepler problem, resulting in an integration error of order $\epsilon \tau^2$.  Thus, the appropriate step size is simply a function of the perturbation magnitude.   For typical solar system integrations, roughly 10-20 steps per orbit are suitable for an accurate symplectic integration \citep{morb02,wishol92}.  The second advantage of symplectic techniques is that the numerical errors for a system's integrals of motion can be expressed in Hamiltonian form \citep{sahtre92, yosh93}.  Hence, for sufficiently small step sizes the errors will not grow with time.  

During a symplectic integration, the numerical Hamiltonian conserves quantities analogous to integrals of motion, but the traditional integrals of motion (e.g. energy) of the numerical system actually oscillate about those of the real Hamiltonian.  Furthermore, this oscillation has a period and amplitude determined by the chosen timestep and Hamiltonian partition \citep{sahtre92}.  \citet{wis96} show that this oscillation takes the form of an infinite series and can be well-represented by its leading terms.  Fortunately, these terms can be solved for and applied as a symplectic corrector at any point during a symplectic integration to reduce numerical error to order $\epsilon^2 \tau^2$.  Thus, when $\epsilon$ is small Wisdom-Holman style symplectic methods have the potential to be much more accurate than their leapfrog form initially implies.  

\begin{figure}[htbp]
\centering
\includegraphics[scale=.85]{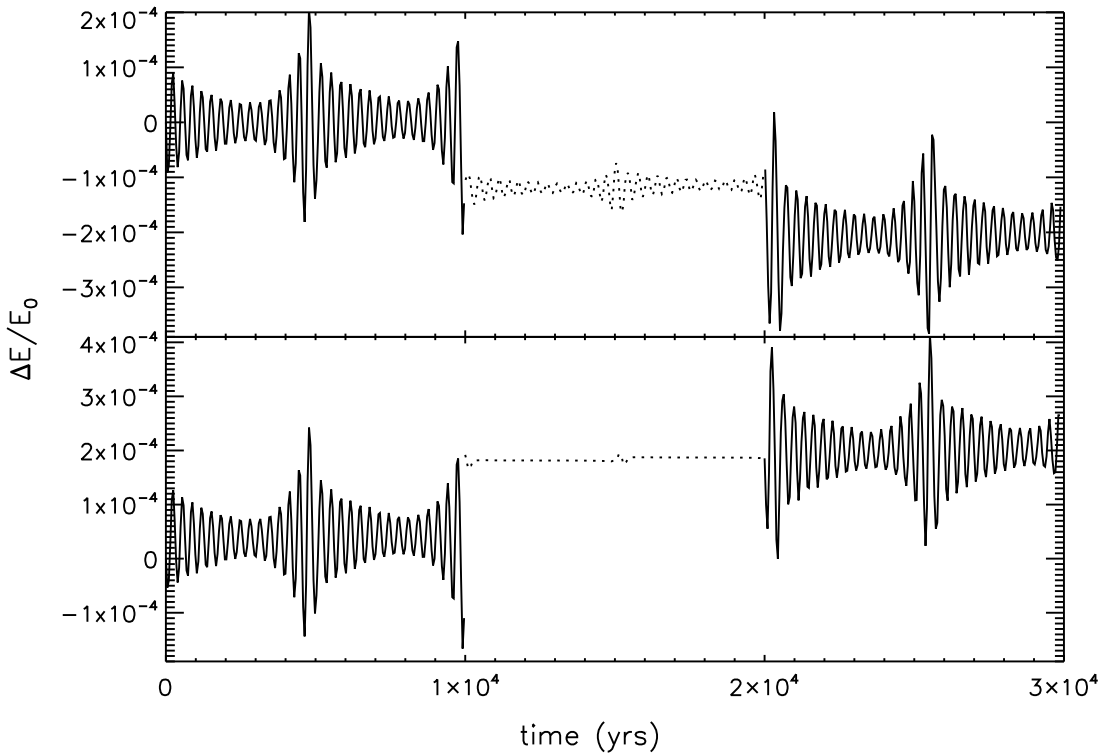}
\caption{Plots of numerical energy error vs. time for an orbit with $a=$300 AU, $q=$100 AU, and $i=$45$^{\circ}$.  {\bf Top Panel:} A heliocentric symplectic integrator is used for this integration.  The solid line corresponds to periods when the integration step was set to 0.548 yrs, and the dotted line marks periods using a 0.27-yr step.  {\bf Bottom Panel:} A symplectic integrator is used for this integration.  The time step is held at 0.548 yrs for the entire integration.  The solid line corresponds to periods when the Keplerian drift is performed about the Sun, and the dotted line corresponds to a Hamiltonian partition using a barycentric drift.}\label{fig:errex}
\end{figure}

As with all methods, symplectic integrators also suffer from a few limitations.  To remain symplectic, several numerical parameters must remain fixed throughout an integration.  One of these is the integration timestep.  The numerical Hamiltonian is a function of the chosen timestep, and thus changing this step will generate a different numerical Hamiltonian.  Unless the integrals of motion of the numerical and real Hamiltonian happen to agree exactly when the timestep is changed, these quantities will begin oscillating about values different from the real system's.  An example of this type of error introduction is shown in the top panel of Figure \ref{fig:errex}, where we integrate a particle on an orbit with semi-major axis $a=$ 300 AU and perihelion distance $q=$ 100 AU for 30,000 yrs in the potential of the Sun and four giant planets.  This particle never comes close to the planetary region.  Consequently, the particle receives only incredibly small energy kicks from the planets during perihelion passages, and the orbital energy variations shown in the figure are almost purely due to numerical error.  Initially, we integrate with a timestep of 0.548 yrs (200 days), and even though the energy oscillations are large, they are also bounded.  At $t=$10,000 yrs we decrease the timestep by half, and although the energy error decreases substantially with the smaller step size, it is clear that the numerical Hamiltonian is now modeling a system different from the original one.  Finally, this same error is demonstrated again when we switch back to the original timestep at $t=$20,000 yrs and one can see that the numerical Hamiltonian drifts even further from the original system.

Another numerical parameter that cannot be altered without compromising an algorithm's symplectic properties is the way the Hamiltonian is partitioned.  The choice of canonical coordinates that are used in a simulation typically determines how the Hamiltonian is split between $H_{Kep}$ and $H_{Int}$.  Changing this splitting will result in an error analogous to a time step change.  Once again, we illustrate this effect in the bottom panel of Figure \ref{fig:errex}:  here we integrate the same particle as in the top panel, but in this case the timestep is held constant for the entire integration.  Initially, the integrator uses a symplectic algorithm that performs Keplerian drifts about the Sun (in heliocentric canonical coordinates).  Then, at $t=$10,000 yrs the Hamiltonian is reexpressed in canonical Jacobian coordinates (which are barycentric for massless test particles).  Now, the drifts are performed about the solar system barycenter.  Because $H_{Int}$ is so much smaller at large distances with this Hamiltonian formulation, the energy oscillations are greatly diminished.  As in the top panel, however, we see that the numerical Hamiltonian is now tracking the behavior of a system that is different from our initial integration.  Furthermore, switching back to the heliocentric algorithm at $t=$20,000 yrs shows that the particle is now on a slightly different orbit than it started with, and additional error is introduced.  

It should be noted that changing either the canonical coordinates or the timestep a handful of times during an integration will not typically invalidate a simulation.  After a single change, the resulting numerical Hamiltonian and error magnitude will still be very close to that of the initial simulation.  In the widely used SWIFT RMVS3 simulation package \citep{levdun94}, both the timestep and canonical coordinate system are changed during encounters between massless test particles and massive bodies.  However, because these encounters are highly chaotic and occur relatively infrequently, the small errors introduced do not alter the actual dynamics of particles \citep{levdun94,mig97}.  However, when timestep and coordinate changes are made many times, each individual transition manifests itself as a step in a random walk away from the true integrals of motion of the system, and in certain situations this random walk can be the driving force behind the dynamics of orbits rather than the real Hamiltonian of the system.

While the above limitations of symplectic routines are inconsequential for many scenarios, there are classes of problems when they impede accurate modeling.  One such class is simulating planetary accretion.  During close encounters between massive bodies in accretion simulations, $H_{Int}$ temporarily dominates $H_{Kep}$, and its treatment as a perturbation to $H_{Kep}$ is no longer valid, which causes the integrator to fail.  Several approaches have been devised to circumvent this problem.  For example, \citet{dun98} decompose planetary forces into a series of forces acting over different ranges and sampled at different rates.  Only during an encounter is it necessary to sample the high-frequency close-range forces since they will be zero when massive bodies are far from one another.  Alternatively, \citet{cham99} temporarily switches to a Bulirsch-Stoer technique to integrate an $H_{Kep}$ expression that absorbs encounter terms when they become too large.  Although this second technique is not strictly symplectic, the Bulirsch-Stoer integration can be performed to machine precision resulting in no additional accuracy loss.  Lastly, \citet{mcnel09} combine these methods with a leapfrog approach assigning different fixed steps to different bodies \citep{satre94} to obtain a quasi-symplectic algorithm that successfully handles encounters and accurately adjusts the integration step for different radial zones.  

Oort Cloud dynamics is another regime hampered by integrator limitations.  When simulating the Oort Cloud, it is necessary to accurately integrate extremely eccentric test particle orbits.  Consequently, a symplectic integrator whose $H_{Kep}$ is centered on the solar system barycenter cannot be used since this approximation fails during very close perihelion passages \citep{rauhol99}.  For this reason, an algorithm such as the SWIFT RMVS3 integrator, which performs Keplerian drifts about the Sun, is typically used instead, e.g. \citet{dones04}.  However, this algorithm is quite inefficient for distant cometary orbits.  Oort Cloud bodies orbit the Sun on Myr-timescales, but the simulation step size must be held below $\sim$0.5 yrs ($\sim5$\% of Jupiter's period) in order to accurately model the perturbations of the giant planets.  This limitation even holds for orbits that do not approach the planetary region because integrations in the heliocentric frame always contain an indirect acceleration term due to planetary perturbations on the Sun, which does not fall off with particle distance.

We present an accurate integration package called SCATR (Symplectically Corrected Adaptive Timestep Routine) that symplectically integrates test particles in a barycentric frame with large timesteps while they are far from the planetary region of the Solar System.  When test bodies approach the planetary region, the integrator switches to heliocentric coordinates and uses a much smaller step size.  To do this smoothly, the algorithm uses a symplectic corrector to drastically reduce the numerical error from time step changes as well as differences in the Hamiltonian partition.  Although the timestep transitions and Hamiltonian reformulations in our algorithm are not symplectic, we show that the net numerical errors accrued by these transitions over the history of the Solar System are still comparable to, or even smaller than, the numerical error of an uncorrected symplectic integration.  The benefit of this additional small numerical error is that the computing time of distant orbital integrations can be decreased by up to a factor of $\sim30$.  While not as elegant as the adaptive step scheme of \citet{mcnel09}, our algorithm is simple and efficient.  As such, it is well-suited for integrations focused purely on test particle dynamics such as Oort Cloud studies.  

\section{The Algorithm}
Our numerical code, which is based on the routines found in SWIFT's RMVS3 \citep{levdun94}, uses the following symplectic integration operator:
\begin{equation}
{\cal E}(\tau) = {\cal I}(\tau/2)\circ {\cal K}(\tau)\circ {\cal I}(\tau/2)
\end{equation}
where the operator ${\cal I}(t)$ corresponds to evolution under the equations of motion of $H_{Int}$ for a time $t$, ${\cal K}$ corresponds to evolution under the equations of motion of $H_{Kep}$, and ${\cal E}(\tau)$ corresponds to evolution of the entire numerical Hamiltonian over one step, $\tau$.  Unlike SWIFT, ours is a hybrid code that only uses the regularized mixed variable symplectic (RMVS) mapping scheme \citep{levdun94} when test particles are close to the planetary region.  When test particles are far from the planetary region our code uses the traditional Wisdom-Holman (WH) symplectic mapping, which performs Keplerian drifts about the barycenter for test particles \citep{wishol91}.  These two mapping schemes differ by the assumed mass about which a particle is orbiting.  For the WH scheme, it is assumed that particles are following nearly Keplerian orbits about the center-of-mass of the solar system.  For test particles under the gravitational influence of the Sun and planets
\begin{equation}
H_{Kep} = \frac{1}{2}v_{b}^2 - \frac{GM}{R}
\end{equation}
in this mapping scheme, where $v_{b}$ is the barycentric particle velocity, $M$ is the solar system mass, and $R$ is the particle distance to the barycenter.  In this scheme the perturbing Hamiltonian is 
\begin{equation}
H_{Int} = \frac{GM}{R}-\frac{GM_S}{R_S}-\displaystyle\sum_{p}\frac{GM_p}{r_p}
\end{equation}
where $M_S$ is the mass of the Sun, $R_S$ is the distance from the particle to the Sun, the $M_p$ terms are the planetary masses, and the $r_p$ terms are the particle-planet distances for each of the planets in the system.  

Alternatively, the RMVS mapping scheme assumes that particles are following nearly Keplerian orbits about the Sun rather than the barycenter.  Consequently, $H_{Kep}$ and $H_{Int}$ take on different forms in this map.  For a test particle, the Keplerian Hamiltonian is now
\begin{equation}
H_{Kep} = \frac{1}{2}v_{S}^2 - \frac{GM_S}{R_S}
\end{equation}
where $v_{S}$ is the heliocentric velocity of the test particle.  For this same particle, the interaction Hamiltonian is now
\begin{equation}
H_{Int} = -\displaystyle\sum_{p}\frac{GM_p}{r_p}-\displaystyle\sum_{p}\frac{GM_p({\bf r_S}-{\bf r_p})}{|{\bf r_S}-{\bf r_p}|^{3}} \cdot {\bf r_S}
\end{equation}
In this expression, the last summation represents the ``indirect'' acceleration experienced by the particle due to the acceleration of the Sun (the RMVS coordinate system origin) by the planets orbiting it.

This map switching in our code is done for two reasons.  First, unlike the WH map, RMVS mapping ensures an accurate integration through perihelion passage for any orbital eccentricity since the Keplerian drift is performed about the Sun.  Second, while $H_{Int}$ for the RMVS method will always contain a large indirect acceleration term due to planetary perturbations on the Sun, the WH map approaches a pure Keplerian drift at large distances since $H_{Int}$ approaches zero there.  If we can devise a method to smoothly switch timesteps and maps (which we detail below), computing efficiency may be greatly enhanced by using much larger integration steps for particles at large distances.

As shown in Figure \ref{fig:errex}, varying the Hamiltonian partition or integration timestep of a symplectic integrator introduces a large amount of error due to spurious oscillations of the system's integrals of motion under the numerical Hamiltonian.  However, symplectic correctors remove a large degree of this oscillation \citep{wis96}, and outfitting our code with a corrector enables us to accurately switch timesteps and symplectic maps far from the Sun.  Following the notation of \citet{wis06}, symplectic correctors can be expressed in terms of ${\cal K}$ and ${\cal I}$.  First, let

\begin{equation}
{\cal X}(a\tau,b\tau) = {\cal K}(a\tau)\circ{\cal I}(b\tau)\circ{\cal K}(-a\tau).
\end{equation}
where $a$ and $b$ are defined coefficients for a corrector of a given order.  Next let
\begin{equation}
{\cal Z}(a\tau,b\tau) = {\cal X}(a\tau,b\tau)\circ {\cal X}(-a\tau,-b\tau).
\end{equation}
Now, an $(n+1)$-order corrector can be written as 
\begin{equation}
{\cal C}(\tau) = {\cal Z}(a_1\tau,b_1\tau)\circ {\cal Z}(a_2\tau,b_2\tau)\circ...\circ{\cal Z}(a_n\tau,b_n\tau).
\end{equation}
In addition, an inverse corrector to convert from the real Hamiltonian to the numerical one can be expressed as
\begin{equation}
{\cal C}^{-1}(\tau) = {\cal Z}(a_n\tau,-b_n\tau)\circ...\circ{\cal Z}(a_2\tau,-b_2\tau)\circ{\cal Z}(a_1\tau,-b_1\tau).
\end{equation}
Thus, a corrected symplectic algorithm that takes $m$ steps will have the following form:
\begin{equation}
{\cal E}'(m\tau) = {\cal C}^{-1}(\tau)\circ \left[{\cal I}(\tau/2)\circ {\cal K}(\tau)\circ {\cal I}(\tau/2)\right]^{m}\circ {\cal C}(\tau).
\end{equation}
or
\begin{equation}
{\cal E}'(m\tau) = {\cal C}^{-1}(\tau)\circ \left[{\cal E}(\tau)\right]^{m}\circ {\cal C}(\tau)
\end{equation}
Note that the corrector and its inverse only need to be applied at the very beginning and at the end of the sequence of $m$ steps since they will cancel each other out in between individual steps.  Finally, a corrected algorithm that takes $m$ steps in the RMVS mapping with step size $\tau_R$ and then switches to the WH mapping before taking $n$ steps of size $\tau_W$ can be written as
\begin{equation}
{\cal E}'(m\tau_R+n\tau_W) = {\cal C}_R^{-1}(\tau_R)\circ\left[{\cal E}_R(\tau_R)\right]^{m}\circ{\cal C}_R(\tau_R)\circ{\cal C}_W^{-1}(\tau_W)\circ\left[{\cal E}_W(\tau_W)\right]^{n}\circ{\cal C}_W(\tau_W)
\end{equation}
where the $R$ subscripts refer to numerical integrations and correctors that perform drifts about the Sun (RMVS map) and the $W$ subscripts mark integrations and corrections performing drifts about the barycenter (WH map).

Now that we have an algorithm defined that combines the advantages of both the RMVS and WH maps, it is necessary to determine where to switch between maps.  As one moves further from the planetary region the solar system's potential approaches that of a monopole and test particle motion will consequently approach Keplerian motion.  At 150 AU, the magnitude of the quadrupole moment is only 10$^{-6}$ of that of the monopole \citep{dun87}.  Through trial and error, we have found the code to behave well if we switch maps at a barycentric distance of 300 AU, where the motion of test particles is very smooth.  (Alternative transition radii are discussed in the next section.)

In regimes of very smooth test particle motion, extensions to higher order approximations will yield large increases in integrator accuracy.  As demonstrated in the error analysis of the next section, we have found that only a third-order (2-stage) symplectic corrector is suitable for our purposes.  This is advantageous because the computational expense of correctors scales directly with stage number.  The coefficients for this third-order corrector are \citep{wis06}
\begin{eqnarray}
\nonumber \lefteqn{a_1 = \frac{3}{10}\gamma,\quad b_1 = \frac{1}{72}\gamma }\\
& & a_2 = \frac{1}{5}\gamma, \quad b_2 = \frac{-1}{24}\gamma
\end{eqnarray}
where $\gamma = 10^{1/2}$.

Simply switching maps and timesteps {\it after} a particle crosses the $r=$ 300 AU boundary will produce an integration that is not time-reversible.  However, \citet{quinn97} show that using a time-reversible step-selection routine significantly reduces the error incurred from varying timesteps in symplectic integrations.  As a result, we use an algorithm that attempts to enforce time-reversibility, i.e. the same sequence of step sizes are used if a particle is integrated in reverse.  To do this, we first  drift all particles about the solar system barycenter for $\tau_W$.  Only particles that have both initial and final positions further than 300 AU are integrated using $\tau_W$ and the WH map.  All other particles are reset to their initial positions and integrated with $\tau_R$ and the RMVS map.  It must be mentioned that even with this algorithm, time-reversibility is still occasionally violated when the end positions of the RMVS, $\tau_R$ integration and the WH, $\tau_W$ drift fall on opposite sides of the $r=300$ AU boundary.  Given the smooth nature of particle motion in this region of the solar system, however, this occurrence is relatively rare, and this algorithm is still a vast improvement over more simplistic routines \citep{quinn97}.

Now that we have outlined our numerical method, we look at its performance in the next section.

\section{Code Performance}

In this section we look at the performance of our method under a few test cases and compare this with other simulation techniques, such as SWIFT's RMVS3 as well as cruder multi-stepping methods. We examine numerical errors in energy and angular momentum, the effect of passing stars and the Galactic tide on distant orbits, and the change in Jacobi constant in the restricted three-body problem.  In addition to this, we investigate how numerical error changes as we vary $\tau_W$ and the step/map transition radius, $r_{trans}$.  Last, we measure how the computation time of our code scales with particle number.  Unless specified otherwise, simulations using SCATR are set to $\tau_R=0.548$ yrs, $\tau_W=90=49.3$ yrs, and $r_{trans}=300$ AU.

\subsection{Numerical Error}
We begin our error analysis with a 1-Gyr integration of a test particle on an orbit with semimajor axis $a=300$ AU, perihelion distance $q=100$ AU, and inclination $i=45^\circ$ under the gravitational influence of the Sun and the four giant planets on their present day orbits.  Since this particular orbit does not approach the planets but does pass through the map/timestep transition radius of our new code twice per orbit, it provides a good measure of the numerical error in Keplerian energy and angular momentum.  Of course, a particle under the influence of the Sun and planets does not actually conserve these quantities, but because this orbit stays far from the planets, these quantities change very little and allow numerical error to be isolated.  

\begin{figure}[htbp]
\centering
\includegraphics[scale=.85]{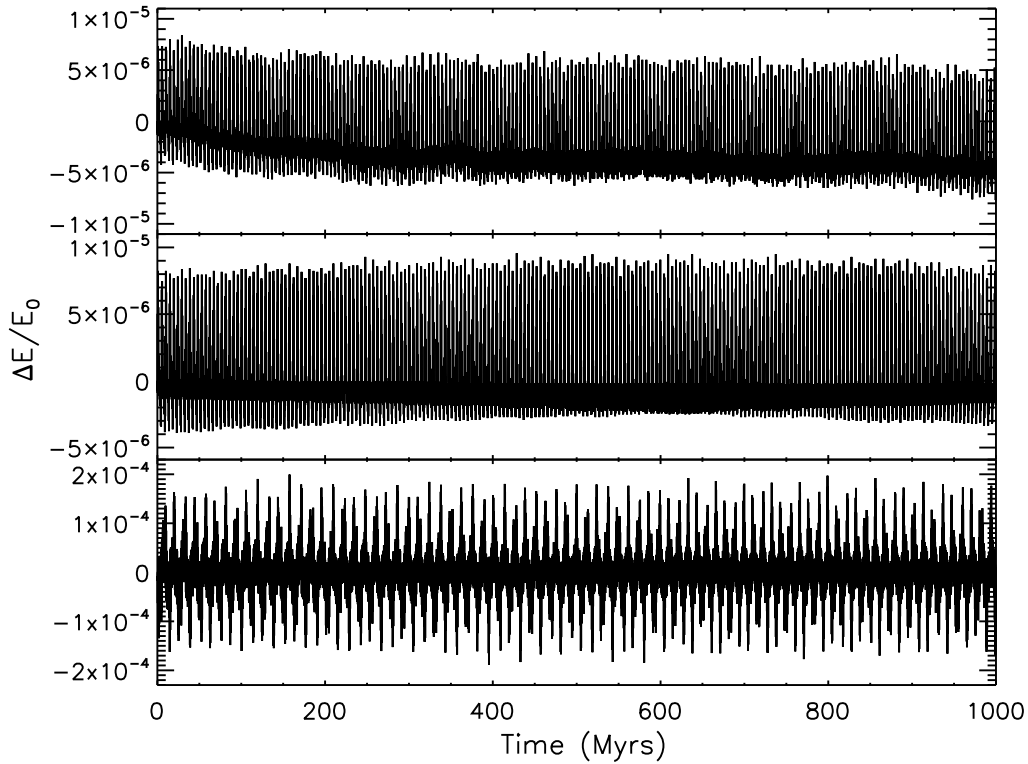}
\caption{Plots of orbital energy error vs. time for 1-Gyr integrations of a test particle orbit with $a=$300 AU, $q=$100 AU, and $i=$45$^{\circ}$.  {\bf Top Panel:}  This integration is performed with our hybrid code, SCATR.  {\bf Middle Panel:}  This integration is performed with a version of SWIFT RMVS3 outfitted with a third-order symplectic corrector.  {\bf Bottom Panel:} This integration is performed with the SWIFT RMVS3 integrator.}\label{fig:longerr}
\end{figure}

In Figure \ref{fig:longerr}, we plot the energy error for three different integrators: our new code (SCATR), a version of SWIFT RMVS3 with a symplectic corrector, and the normal version of SWIFT RMVS3.  We see that after 1 billion years the integration using our new code has an energy error of less than one part in 10$^{5}$.  This error is about 30\% larger than that of a symplectically corrected integration done entirely in the RMVS map with a constant timestep of 0.548 yrs.  However, most solar system symplectic integrations do not use a corrector, and when we compare the error of SCATR with a normal RMVS3 integration we see that although ours shows some secular drift, our error is over an order of magnitude smaller than the uncorrected routine. 

\begin{figure}[htbp]
\centering
\includegraphics[scale=.85]{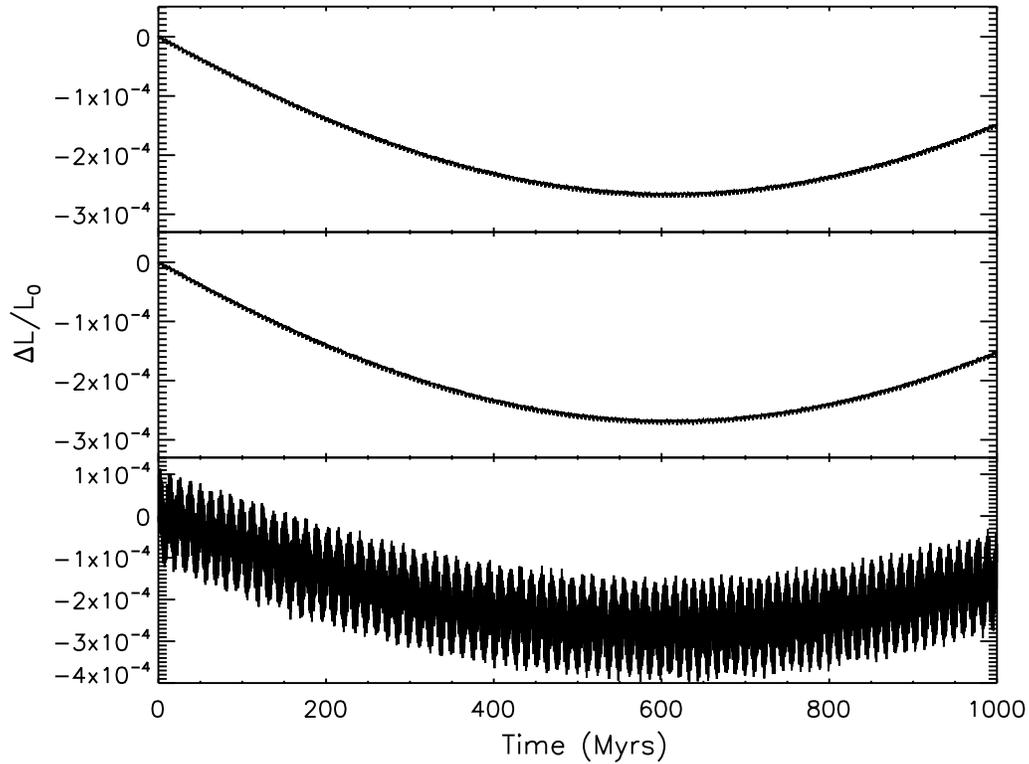}
\caption{Plots of orbital angular momentum vs. time for 1-Gyr integrations of an orbit with $a=$300 AU, $q=$100 AU, and $i=$45$^{\circ}$.  {\bf Top Panel:}  This integration is performed with SCATR.  {\bf Middle Panel:}  This integration is performed with a version of SWIFT RMVS3 outfitted with a third-order symplectic corrector.  {\bf Bottom Panel:} This integration is performed with the SWIFT RMVS3 integrator.}\label{fig:longerr2}
\end{figure}

An examination of orbital angular momentum evolution for the three integrations yields similar results.  The change in orbital angular momentum is plotted for each of the three simulations in Figure \ref{fig:longerr2}.  Unlike the energy, angular momentum evolves noticeably during our test particle integration.  The reason for this is that the higher multipole moments of the solar system's potential produce a small secular change in the angular momentum, or $L$, over time.  As can be seen in the upper two panels of Figure \ref{fig:longerr2}, the angular momentum evolution of the SCATR simulation is nearly identical to the RMVS version that includes a corrector.  Even though the solar system potential closely resembles a monopole at distances beyond $\sim$150 AU, the orbital evolution in our hybrid simulation is much more sensitive to the weak effects of the higher order multipoles of the solar system rather than the secular effects caused by numerical error, which is encouraging.  Lastly, the bottom panel of Figure \ref{fig:longerr2} demonstrates that the normal version of SWIFT undergoes the same evolution in $L$, but experiences spurious oscillations that are an order of magnitude greater than when a corrector is employed.

It should be noted that with such a large $\tau_W$ ($\sim$50 yrs) our sampling frequency of the giant planet perturbations is less than some of their orbital frequencies.  However, the frequency of a perturbation due to a single orbit of a planet is extremely high (as well as small in magnitude) compared to the orbital periods of test particles being integrated with $\tau_W$.  Consequently, due to the averaging principle, these weak, high frequency perturbations should not affect the long-term evolution of distant test particles.  Moreover, the smooth change in $L$ and its match to a fixed-step integration in Figure \ref{fig:longerr2} demonstrates that our $\tau_W$ sampling is still capturing the long-term deviations of the solar system potential from a monopole (the higher multipole moments) that are critical to the evolution of distant orbits.  (We tested earlier versions of our code that instead modeled the distant solar system potential as just a monopole or a monopole with fixed $J_2$ and $J_4$ terms.  However, this resulted in the systematic inclusion and exclusion of higher multipole terms at specific test particle distances.  This produced systematic errors that accumulated quickly for some test particle orbital inclinations where these multipole terms significantly impacted orbital evolution.)

\begin{figure}[htbp]
\centering
\includegraphics[scale=.85]{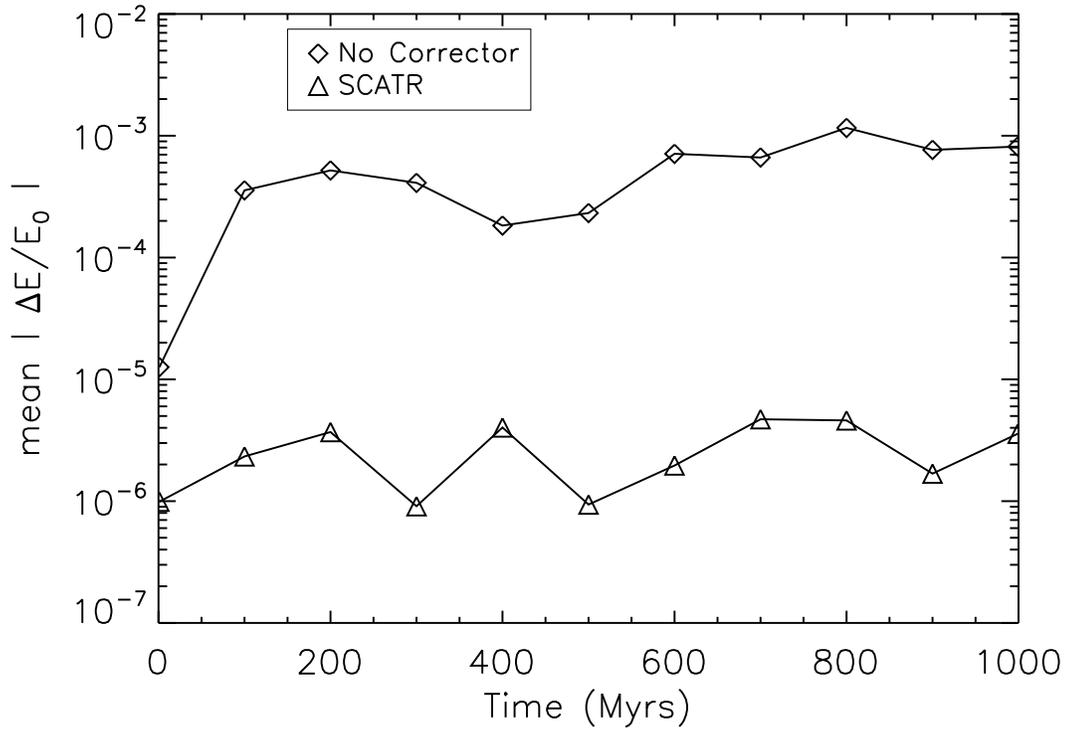}
\caption{A comparison of the numerical energy error vs. time for an integration using SCATR ({\it triangles}) and an integration where the use of a symplectic corrector during timestep changes is disabled ({\it diamonds})  These integrations include the Sun and four giant planets as well as 20 test particles.  The mean energy error of each suite of particles is plotted.}\label{fig:crude}
\end{figure}

Figures \ref{fig:longerr} and \ref{fig:longerr2} illustrate that the numerical error of SCATR is comparable to widely used simulation methods over Gyr timescales.  However, we have yet to demonstrate how vital the symplectic corrector is to obtaining the small numerical errors seen in the previous two figures.  Figure \ref{fig:crude} illustrates this by comparing the energy error growth of SCATR with that of an integration in which a symplectic corrector is not used during timestep transitions.  In this experiment, we use the two different integrators to integrate the orbits of twenty test particles for 1 Gyr.  The perihelia and semimajor axes of these orbits are set to 100 AU and 250 AU respectively to isolate the numerical energy error and produce the highest number of time step transitions per particle (inclinations were chosen randomly).  In Figure \ref{fig:crude}, we plot the growth of the mean numerical energy error for our suite of particles for each of the algorithms.  As can be seen in the plot, if the use of a symplectic corrector is excluded during timestep transitions, the energy error increases by 2--3 orders of magnitude.  Thus, it is critical to use a symplectic corrector if we wish to minimize the numerical error due to timestep transitions in our algorithm.

\begin{figure}[htbp]
\centering
\includegraphics[scale=.85]{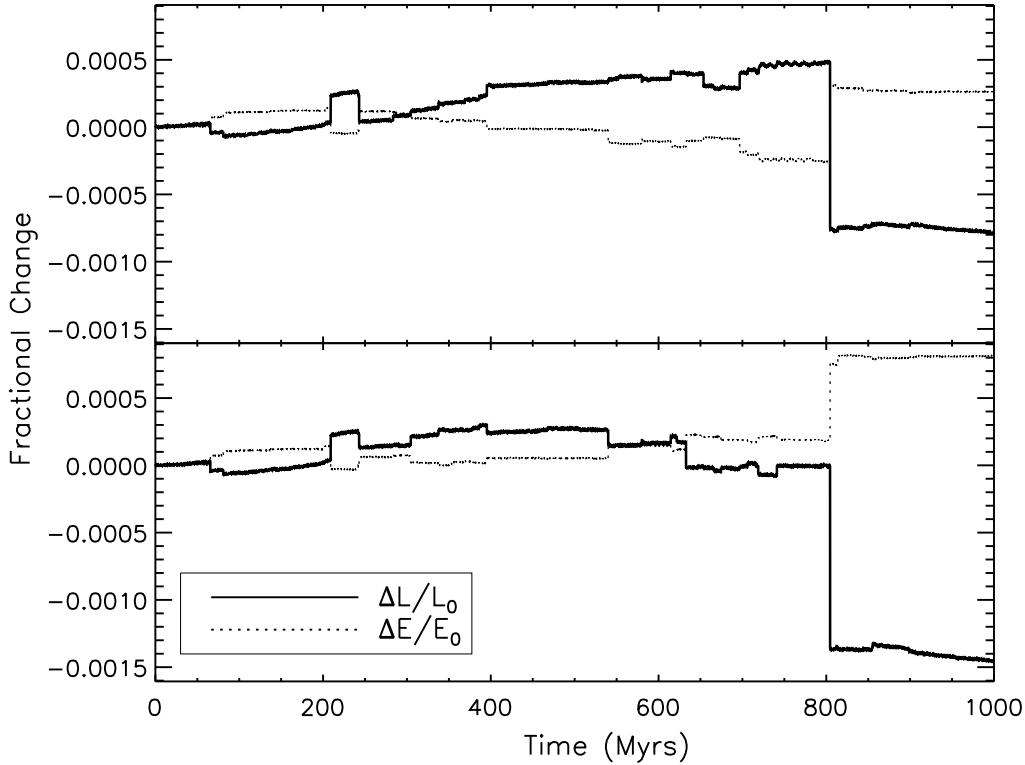}
\caption{Plots of fractional change in orbital energy ({\it dotted line}) and angular momentum ({\it solid line}) vs. time for an orbit with $a=$300 AU, $q=$100 AU, and $i=$45$^{\circ}$.  These integrations include the effects of passing stars and the Milky Way tide.  {\bf Top Panel:}  This integration is performed with SCATR.  {\bf Bottom Panel:} This integration is done with a version of the SWIFT RMVS3 integrator that includes a symplectic corrector.}\label{fig:errpert}
\end{figure}

To isolate the numerical error of our integrator from other effects we have purposely not included other gravitational perturbations that become important at large distances from the Sun, such as passing stars and the galactic tide.  Even at orbital distances of a few hundred AU, both stellar passages and the galactic tide will drive changes in orbital energy and angular momentum that are much larger than those caused by our numerical error.  When we rerun our previous integrations and include a population of passing stars \citep{rick08} and a galactic tidal model \citep{lev01}, we see in Figure \ref{fig:errpert} that the energy drifts in each simulation are much greater than before, and are caused by stochastic impulses from passing stars.  Figure \ref{fig:errpert} also shows that the changes in orbital angular momentum are now much larger than previously due to both tidal and stellar perturbations.  Lastly, the fact that our two integrations experience $E$ and $L$ drifts of similar magnitude but finish with different endstates indicates that chaotic evolution under passing stars and the galactic tide causes slightly different orbits to diverge on Gyr timescales.  When these effects are considered, the small errors due to timestep and coordinate changes become even less significant.

Our code was conceived in order to increase the efficiency of Oort Cloud integrations, and because forming the Oort Cloud involves repeated scatterings by giant planets, another important test is to measure how well our code handles close encounters between test particles and massive bodies.  To evaluate this, we take the case of the restricted 3-body problem.  In this simulation, Neptune is placed on a circular orbit around the Sun at 30 AU and a test particle on an orbit with semimajor axis $a=200$ AU, perihelion distance $q=27$ AU, and inclination $i=10^\circ$.  This system is then integrated for 100 Myrs.  This particular orbit configuration was chosen because it forces the particle to interact strongly with Neptune as well as cross the timestep/coordinate transition boundary at $r=300$ AU twice per orbit.

\begin{figure}[htbp]
\centering
\includegraphics[scale=.85]{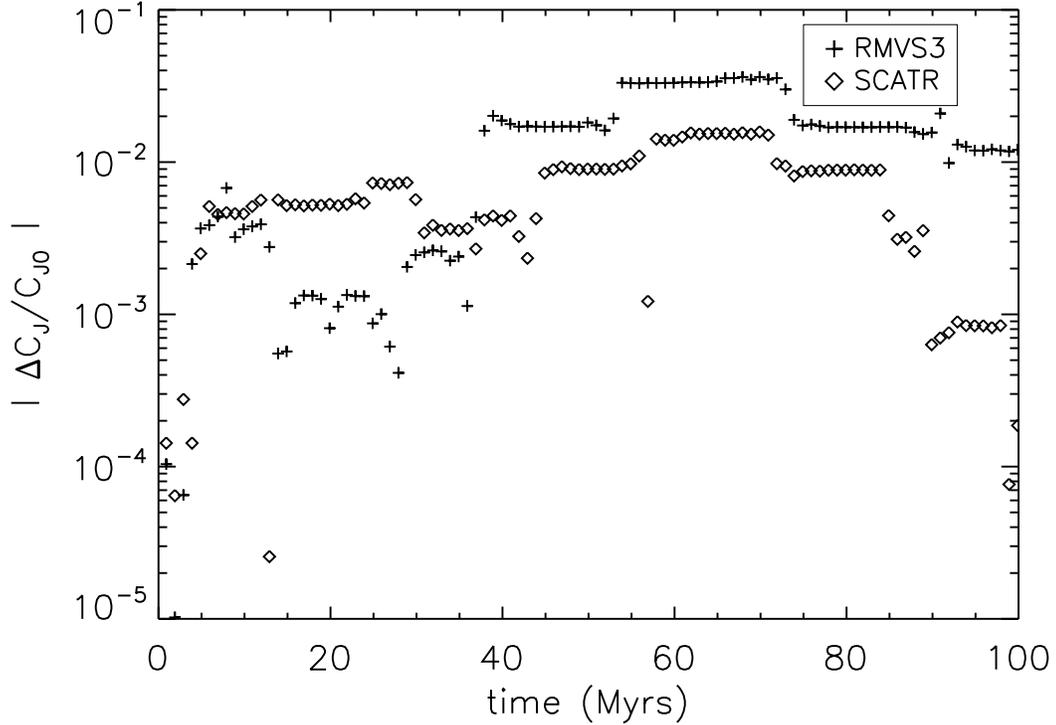}
\caption{Plot of fractional change in Jacobi Constant vs. time for two integrations of a test particle under the gravitational influence of a circularly orbiting Neptune mass planet at 30 AU.  The first integration was performed with the SWIFT RMVS3 package ({\it crosses}), and the second integration was performed with SCATR ({\it diamonds}).}\label{fig:errjac}
\end{figure}

Two integrations of this system are performed: one using the RMVS3 version of SWIFT and the other using SCATR.  The results of these integrations are shown in Figure \ref{fig:errjac} where we plot the drift in the Jacobi constant of the test particle with time for each of our integrations.  This figure demonstrates that both integration packages experience similar levels of numerical error in terms of change in the Jacobi constant.  The reason for this similar behavior is that the additional error introduced from our timestep and Hamiltonian partition change at $r=300$ AU is negligible compared to the numerical error introduced by close encounters.  Thus, our code seems to integrate planetary encounters as well as the SWIFT integration package.

\begin{figure}[htbp]
\centering
\includegraphics[scale=.85]{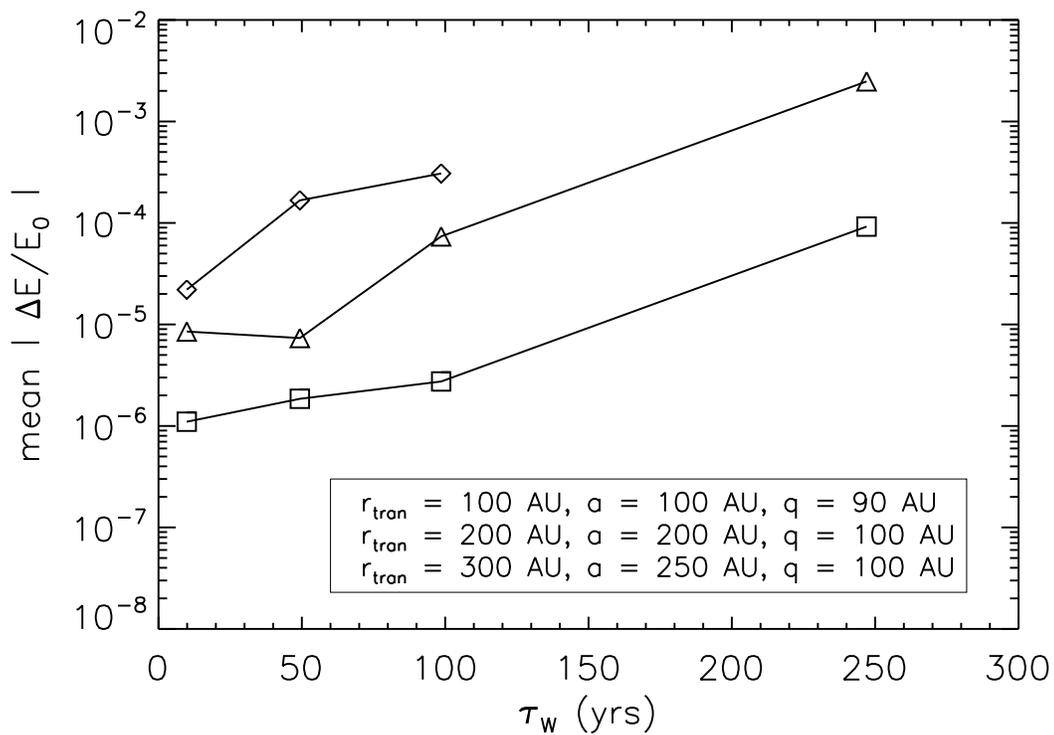}
\caption{Plot of mean energy error vs. $\tau_W$ for 11 different 1-Gyr simulations of 20 test particles orbits.  $r_{trans}$ was varied between 100 ({\it diamonds}), 200 ({\it triangles}), and 300 AU ({\it squares}).  The test particle orbits for each $r_{trans}$ value are listed in the figure legend.  Simulations included the Sun and four giant planets.}\label{fig:errcomp}
\end{figure}

We have now established that the numerical errors of SCATR are smaller than or comparable to that of SWIFT's RMVS3 for solar system simulations that use $\tau_W = 49.3$ yrs and $r_{trans} = 300$ AU.  However, different degrees of accuracy and computing efficiency can be attained if these two parameters are adjusted.  For this reason, we now measure how the numerical error of our code changes as we vary $\tau_W$ and $r_{trans}$.  To do this, we run 11 different 1-Gyr simulations, each one using a different combination of $\tau_W$ and $r_{trans}$ to integrate 20 test particles in the presence of the Sun and giant planets.  To set up these simulations, we first choose $\tau_W$ and $r_{trans}$.  The different values we use for $\tau_W$ are 9.86, 49.3, 98.6, and 247 yrs, while the chosen values of $r_{trans}$ are 100, 200, and 300 AU.  (These particular timesteps are chosen because for SCATR to function properly $\tau_W$ must be an integer multiple of $\tau_R$, which we hold fixed at 0.548 yrs.)  With $\tau_R$ and $r_{trans}$ chosen, we next assign the same perihelion distance and semimajor axis to all particles in a given simulation to model a ``worst-case'' orbit - one with a short orbital period that crosses our chosen $r_{trans}$ but stays far enough from the giant planets to isolate the numerical energy error.  The perihelion and semimajor axis used for each $r_{trans}$ can be found in the legend of Figure \ref{fig:errcomp}.  Finally, we randomly select the remaining orbital elements of the test particles (inclination, longitude of ascending node, argument of perihelion, and mean anomaly).  

The results of our 11 simulations are shown in Figure \ref{fig:errcomp}.  For each simulation we plot the mean energy error per particle as a function of $r_{trans}$ and $\tau_W$.  One can see from this figure that for a given timestep the numerical error of the integrator grows by over an order of magnitude as $r_{trans}$ is moved from 300 AU to 100 AU.  This is for two different reasons.  First, relative to $H_{Kep}$, $H_{Int}$ and therefore $\epsilon$ are larger at 100 AU than at 300 AU.  Consequently, the accuracy gain of the symplectic corrector is smaller at 100 AU since its error scales with $\epsilon^2$.  Second, setting $r_{trans}=100$ AU enables particles with smaller semimajor axes to cross the timestep/map transition than when $r_{trans}$ is set to 300 AU.  Since these particles have shorter orbital periods, they will accrue numerical error faster than their $r_{trans}=300$ AU counterparts.  

The second feature one notices in Figure \ref{fig:errcomp} is the general deterioration of integration accuracy with increasing $\tau_W$.  Again, there are two reasons for this.  As stated previously, the error in a symplectic corrector is of order $\epsilon^2\tau^2$, so the error introduced each time ${\cal C}_W$ is calculated will increase with $\tau_W$.  In addition, the planetary perturbations to pure Keplerian motion are very tiny but still non-zero outside of $r_{trans}$.  While perhaps not important compared with the effects of tidal and stellar perturbations at these distances, the planetary forces are nevertheless being undersampled by $\tau_W$.  As a result, the error due to $H_{Int}$ will not be bounded as in a typical symplectic integration.  This will produce an additional small secular drift in orbital energy.  (Note that we did not run a $\tau_W=247$ yrs, $r_{trans}=100$ AU case since $\tau_W$ would be 1/4 of the test particle orbital periods.)

With the numerical error of our code characterized, we now move on to analyzing its efficiency.

\subsection{Computing Time}

In this section we measure the reduction in computing time attained by our code.  To do this we integrate a sample of ``generic'' Oort Cloud comets, since this is the orbital regime in which we expect our code to perform best.  Our chosen comet orbits have semimajor axis $a$ between 10$^3$ AU and 10$^5$ and a spatial density which scales as $a^{-7/2}$.  In addition, the comets are given isotropic distributions in the other orbital elements.  The comets are then integrated under the gravitational influence of the Sun and the four giant planets on their present day orbits.  Each integration is performed for 1 Myr on an Intel Core 2 duo 2.33 GHz processor twice: once using SCATR (with $\tau_W=49.3$ years and $r_{trans}=300$ AU) and once with an unaltererd version of the SWIFT RMVS3 package from which our code was derived.  

\begin{figure}[htbp]
\centering
\includegraphics[scale=.85]{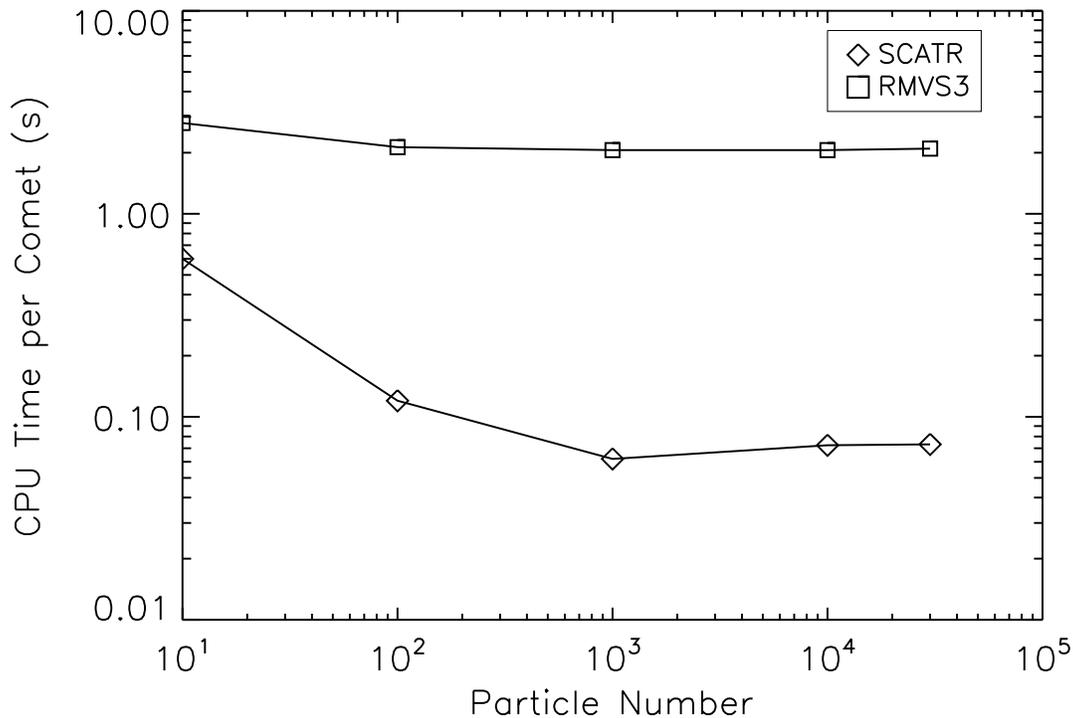}
\caption{Plot of computing time per comet for 1-Myr integrations of Oort Cloud comets vs. total number of comets in a simulation.  Two sets of integrations are performed: one using the SWIFT RMVS3 integrator ({\it squares}) and one using our new integrator, SCATR ({\it diamonds}).}\label{fig:perf}
\end{figure}

The results of our test integrations are shown in Figure \ref{fig:perf} where we plot computing time per comet vs. the number of comets in each simulation.  As can be seen in this plot, our code is significantly faster than RMVS3 for all comet numbers that were tested.  In fact, for simulations with more than 1000 comets the computing time of our code is more than a factor of 30 shorter!  The reason that this speed gain depends on the number of comets is that in SCATR simulations with small numbers of comets, most of the computing time is spent integrating the orbits of the giant planets.  Since high semimajor axis orbital integrations are so computationally cheap in SCATR, the total computing time does not initially change much as the number of comets increases, causing the time per comet to fall quickly at first.  Only above $N \gtrsim$ 1000, when the computing load is dominated by comet orbit integrations rather than planetary ones, does computing time per comet reach a constant value.  On the other hand, test particles and planets are always integrated with the same time step in RMVS3 simulations, making their computational expense more comparable to one another.  Consequently, the computing time per comet is essentially constant for all of our RMVS3 simulations.

\section{Summary}

The computing time of orbital integrations that transition between regimes of vastly different dynamical timescales can be greatly reduced by using different integration step sizes and canonical coordinate systems for each regime.  However, repeatedly changing these two numerical parameters in a symplectic algorithm introduces an accumulation of numerical error that can alter the dynamics of the system being modeled.  We have shown that symplectic correctors can be used to eliminate the bulk of the error introduced during timestep and coordinate changes.  Implementation of this technique yields an integrator that is dramatically faster for simulations of distant test particle orbits.  Additionally, it has energy and angular momentum errors comparable to normal symplectic routines, such as SWIFT's RMVS3.  By decreasing the computing expense of distant orbit integrations by up to a factor of $\sim$30, our code has the potential to greatly increase the statistical significance of many outer solar system simulations.  We will make our code available to any interested users, and they should contact us via email to obtain a copy.  

\section{Acknowledgements}

This work was funded by a NASA Earth and Space Science Fellowship as well as a grant from the NSF (AST-0709191).  We would like to thank Martin Duncan and Seppo Mikkola for reviewing an earlier version of this work and providing helpful comments.

\bibliography{SymplecticCorrector}

\end{document}